\documentclass[sigconf,screen]{acmart}

\usepackage{siunitx}
\usepackage{multirow}

\usepackage{enumerate}
\usepackage{booktabs}
\usepackage{calc}
\usepackage{paralist}
\usepackage[skip=0pt]{caption}
\usepackage[skip=0pt,labelfont=bf,textfont=bf,singlelinecheck=on]{subcaption}
\usepackage{tikz}
\usetikzlibrary{fit,positioning}
\usepackage{graphicx}
\usepackage{color}
\usepackage{colortbl}
\usepackage{mathtools}
\usepackage{algorithm}
\usepackage{algorithmic}
\usepackage[normalem]{ulem}
\usepackage[inline]{enumitem}
\usepackage{numprint}
\npthousandsep{,}
\npdecimalsign{.}
\usepackage{acronym}
\usepackage{setspace}

\copyrightyear{2020}
\acmYear{2020}
\setcopyright{acmlicensed}
\acmConference{SIGIR '20}{July 25--30, 2020}{Xi'an, China}
\acmPrice{}

\newcommand{\duwen}{\phantom{$^\blacktriangle$}}
\newcommand{\noop}[1]{}

\newcommand{\softmax}{\operatorname*{softmax}}
\newcommand{\sigmoid}{\operatorname*{sigmoid}}

\allowdisplaybreaks

\author{Zhiqiang Pan}
\orcid{}
\affiliation{%
	\institution{Science and Technology on Information Systems Engineering Laboratory\\ National University of Defense Technology}
	\city{Changsha}
	\country{China}
}
\email{panzhiqiang@nudt.edu.cn}

\author{Fei Cai}
\orcid{}
\authornote{Corresponding author.}
\affiliation{%
	\institution{Science and Technology on Information Systems Engineering Laboratory\\ National University of Defense Technology}
	\city{Changsha}
	\country{China}
}
\email{caifei@nudt.edu.cn}

\author{Yanxiang Ling}
\orcid{}
\affiliation{%
	\institution{Science and Technology on Information Systems Engineering Laboratory\\ National University of Defense Technology}
	\city{Changsha}
	\country{China}
}
\email{lingyanxiang@nudt.edu.cn}

\author{Maarten de Rijke}
\orcid{0000-0002-1086-0202}
\affiliation{%
	\institution{University of Amsterdam}
	\city{Amsterdam}
	\city{The Netherlands}
}
\email{derijke@uva.nl}

\title{Rethinking Item Importance in Session-based Recommendation}

\begin{document}

\begin{abstract}
Session-based recommendation aims to predict users' based on anonymous sessions. 
Previous work mainly focuses on the transition relationship between items 
during an ongoing session. 
They generally fail to pay enough attention to the importance of the items 
in terms of their relevance to user's main intent.
In this paper, we propose a Session-based Recommendation approach with an Importance Extraction Module, i.e., SR-IEM, that considers both a user's long-term and recent behavior in an ongoing session. 
We employ a modified self-attention mechanism to estimate item importance in a session, which is then used to predict user's long-term preference. 
Item recommendations are produced by combining the user's long-term preference and current interest as conveyed by the last interacted item. 
Experiments conducted on two benchmark datasets validates that SR-IEM outperforms the start-of-the-art 
in terms of Recall and MRR and has a reduced computational complexity. 
\end{abstract}



\maketitle

\section{Introduction}
\label{Introduction}
Recommender systems help connect people to personalized information in a growing volume of items on offer.
Most existing approaches for recommendation focus on a user's interaction history in order to predict their preferences for recommending future items.
For cases where historical user-item interactions are unavailable, it is challenging to capture the user's preferences in an accurate manner~\citep{ICLR16/GRU}. 
For the task of session-based recommendations, we aim to generate recommendations merely based on an ongoing session.

RNNs, attention mechanisms, and GNNs have been widely applied to session-based recommendation. 
For instance, \citet{ICLR16/GRU} apply a Gated Recurrent Unit (GRU) to model user's sequential behavior in session to capture his instant preference, and \citet{CIKM17/NARM} propose 
to capture user's main purpose with an attention mechanism.
On the basis of NARM, \citet{SIGIR19/CSRM} introduce neighbor sessions as auxiliary information to model an ongoing session. 
In addition, \citet{KDD18/STAMP} estimate user's general and current interests based on a long-term and short-term memory, respectively. 
\citet{AAAI19/SRGNN} employ Gated Graph Neural Networks (GGNNs) to  
model the complex transitions between items for producing predictions.

Even though the approaches listed above have all helped to improve the performance of session-based recommendation, they fail to pay enough attention to an important source of information.
That is, they can not accurately locate the important items in a session for generating user preferences.
After generating item embeddings, the importance of each item is simply determined by its relevance either to the mixture of items in the long-term history \citep{CIKM17/NARM,SIGIR19/CSRM} or the last single item \citep{AAAI19/SRGNN} or a combination \citep{KDD18/STAMP}.
Unavoidably, there are non-relevant items in a session, especially in long sessions, making it hard for models to focus on the important items. 

We propose an approach for \emph{session-based recommendation with an importance extraction module}, i.e., \textbf{SR-IEM}, that can effectively capture a user's long-term preferences and their current interest. 
To model a  user's long-term preference, we propose an Importance Extraction Module (IEM) that applies a modified self-attention mechanism to extract the importance of each item in an ongoing session. 
Then, the items are discriminatively combined to predict a user's general preference according to the item importance. 
To capture a user's current interest, we regard the last item embeddings as an expression of their current interest, which is then combined with the long-term preferences for item recommendation. 

Our contributions in this paper are: 
\begin{enumerate*}
	\item We propose an Importance Extraction Module (IEM) to accurately obtain the importance of each item for session-based recommendation. The proposed SR-IEM model can simultaneously capture a user's long-term preference and his current interest to make recommendations. 
	\item We compare the performance of SR-IEM against start-of-the-art baselines on two public datasets and find that it can beat the state-of-the-art models in terms of Recall and MRR. In addition, SR-IEM has a lower computational complexity than competitive neural baselines.
\end{enumerate*}

\vspace{-3pt}
\section{Approach}
\label{Approach}
\begin{figure}[t]
	\centering
	\includegraphics[width=0.8\columnwidth]{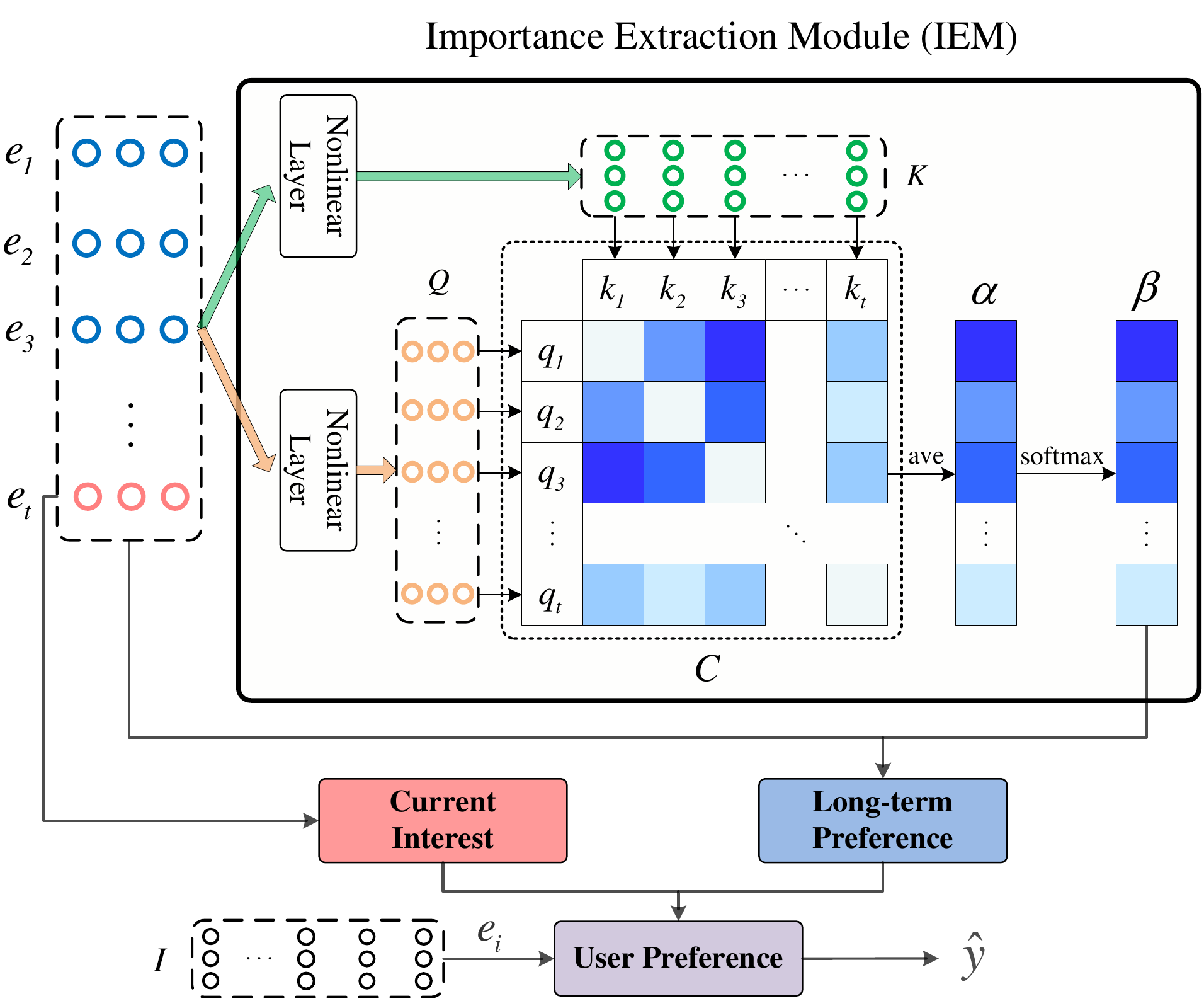}
	\caption{Overview of the proposed SR-IEM model.}
	\label{Figure1}
	\vspace{-5pt}
\end{figure}

Given a session $S = \{x_1, x_2, \ldots, x_t\}$ consisting of $t$ items that a user interacted with, e.g., clicked and purchased, the goal of session-based recommendation is to predict the next item from a set of $n$ items $I = \{v_1, v_2, \ldots, v_n\}$ to recommend at time step $t$+1.
%
%
Fig.~\ref{Figure1} presents an overview of SR-IEM, with three main components, i.e., an importance extraction module (see \S\ref{Importance extraction module}), a preference fusion module (see \S\ref{Preference fusion}), and an item recommendation module (see \S\ref{Prediction}). 

\vspace{-5pt}
\subsection{Importance extraction}
\label{Importance extraction module}

To accurately locate the important items in a session for the purpose of modeling user preference, we propose an Importance Extraction Module (IEM) to generate the item importance. 
We first embed each item $x_i$ in $S = \{x_1, x_2, \ldots, x_t\}$ into a $d$ dimensional representation $e_i \in \mathbb{R}^d$ via an embedding layer. 
Then, borrowing the merits from the self-attention mechanism \citep{NIPS17/Attention}, we transform the item embeddings $E = \{e_1, e_2,\ldots, e_t\}$ into a different space via a non-linear function to generate the respective $\mathit{query}$ $Q$ and $\mathit{key}$ $K$ as:
\begin{align}
\label{equation1}
Q & = \sigmoid(W_q E),
\\
\label{equation2}
K & = \sigmoid(W_k E),
\end{align}
where $W_q \in \mathbb{R}^{d \times l}$ and $W_k \in \mathbb{R}^{d \times l}$ are trainable parameters for the $\mathit{query}$ and $\mathit{key}$, respectively; $l$ is the dimension of the attention mechanism; $\sigmoid$ is the transformation function to learn information from the item embedding in a non-linear way.

After generating the representation of $\mathit{query}$ $Q$ and $\mathit{key}$ $K$, we estimate the importance of each item via a modified self-attention mechanism. 
First, we compute the similarity of every pair of two items by introducing the affinity matrix $C$ between $\mathit{query}$ $Q$ and $\mathit{key}$ $K$ as:
\begin{equation}
\label{equation3}
C = \frac{\sigmoid(Q K^{\mathsf{T}})}{\sqrt{d}},
\end{equation}
where $\sqrt{d}$ is used to scale the attention.

From the affinity matrix, we would like to see that an item is not important if its corresponding similarity scores related to other items are relatively low. 
A user might interact with such an item by accident or due to curiosity. 
In contrast, if an item is similar to most items in a session, it may express the user's main preference. 
That is, the item is relatively important.
Inspired by this intuition, we resort to the average similarity between an item and other items in session as the item importance. 
To avoid high matching scores between identical vectors of $\mathit{query}$ $Q$ and $\mathit{key}$ $K$, following \citep{Arxiv18/AttRec}, we employ a masking operation that masks the diagonal of the affinity matrix. 
Then, we can assign an \emph{importance score} $\alpha_i$ to each item $i$:
\begin{equation}
\label{equation4}
\alpha_i = \frac{1}{t} \sum_{j=1, j \neq i}^t C_{ij}, 
\end{equation}
where $C_{ij} \in C$. 
To normalize the scores, a softmax layer is applied to get the final \emph{importance} $\beta$ of items in the session as:
%
%
%
\begin{equation}
\label{equation5}
\beta = \softmax{(\alpha)},
\end{equation}

\vspace{-5pt}
\subsection{Preference fusion}
\label{Preference fusion}
Through the importance extraction module, we obtain the importance of each item in a session, which indicates the relevance of each item to the user's main purpose. 
Then, we represent the user's long-term preference $z_l$ by combining the embeddings of items in the session according to their importance as:
\begin{equation}
\label{equation6}
z_l = \sum_{i=1}^t {\beta_i e_i}.
\end{equation}
As for the current interest, denoted as $z_s$, following \citep{KDD18/STAMP,AAAI19/SRGNN}, we directly adopt the embedding of the last item in the session, i.e., $z_s = e_t$. 
%
%
After obtaining a user's long-term preference $z_l$ and his current interest $z_s$, we combine them into the user's final preference representation $z_h$ that is to be used for item recommendation as:
\begin{equation}
\label{equation8}
z_h = W_0 [z_l; z_s],
\end{equation}
where $[\cdot]$ is the concatenating operation. $W_0 \in \mathbb{R}^{d \times 2d}$  transforms the concatenated representation from a latent space $\mathbb{R}^{2d}$ into $\mathbb{R}^{d}$.

\vspace{-5pt}
\subsection{Item recommendation}
\label{Prediction}
Once the user's preference representation in a session has been generated, we use it to produce item recommendations by calculating the probabilities for all items in the candidate item set $I$. 
We first compute the user's preference score $\hat {z_i}$ for each item $v_i$ in the candidate item set $I$ by a multiplication operation as:
\begin{equation}
\label{equation9}
\hat {z_i} = z_h^{\mathsf{T}} e_i,
\end{equation}
where $z_h$ is obtained by Eq.~\eqref{equation8} and $e_i$ is the embedding of item $v_i$. 
Then a softmax layer is applied to the preference scores to generate a normalized probability
of each item to be recommended as:
\begin{equation}
\label{equation10}
\hat {y} = \softmax{(\hat {z})},
\end{equation}
where $\hat {z}=(\hat {z}_1, \hat {z}_2, \ldots, \hat {z}_n)$. 
Finally, the items with the highest scores in $\hat {y}$ will be recommended to the user. 

To train our model, we employ the cross-entropy as the optimization objective to learn the parameters as:
\begin{equation}
\label{equation11}
L(\hat{y}) = -\sum_{i=1}^{n}y_i \log(\hat{y}_i) + (1-y_i) \log(1-\hat{y}_i),
\end{equation}
where $y_i\in y$ reflects the appearance of an item in the one-hot encoding vector of the ground truth,
i.e., $y_i = 1$ if the $i$-th item is the target item; 
otherwise, $y_i = 0$. 
Finally, we apply the Back-Propagation Through Time algorithm to train SR-IEM.

\section{Experiments}
\label{Experiments}

\emph{Research questions.}
(RQ1)~Can the proposed SR-IEM model beat the competitive baselines?
(RQ2)~How does SR-IEM perform compared to the baselines under various session lengths?
(RQ3)~How does IEM perform on distinguishing the importance of items in a session compared to other importance extraction methods?

\smallskip\noindent\emph{Model summaries.}
We answer our research questions by comparing the performance of SR-IEM against eight  baselines for session-based recommendation: 
\begin{enumerate*}
	\item Three traditional methods, i.e., S-POP, 
	Item-KNN~\citep{WWW01/ItemKNN} and FPMC~\citep{WWW10/FPMC}; 
	\item Five neural models, i.e., GRU4REC \citep{ICLR16/GRU}, NARM~\citep{CIKM17/NARM}, STAMP~\citep{KDD18/STAMP}, CSRM~\citep{SIGIR19/CSRM} and SR-GNN~\citep{AAAI19/SRGNN}.
\end{enumerate*}

\smallskip\noindent\emph{Datasets and parameters.}
The datasets we use for evaluation are two public benchmark e-commerce datasets, i.e., YOOCHOOSE\footnote{http://2015.recsyschallenge.com/challege.html} and DIGINETICA.\footnote{http://cikm2016.cs.iupui.edu/cikm-cup}
We use the same preprocessing of the datasets as in \citep{CIKM17/NARM,KDD18/STAMP,AAAI19/SRGNN}. 
The statistics of the datasets are listed in Table~\ref{Table1}.
Following \citep{ICDM18/SASRec}, 
we set the maximum session length $L$ to  
10, indicating that for long sessions, we only consider the 10 most recent items. 
\begin{table}[t]
	\centering
	\caption{\textbf{Statistics of the datasets used in our experiments.}}
	\setlength{\tabcolsep}{0.025\linewidth}{
		\begin{tabular}{l r@{}l r@{}l}
			\toprule
			Statistics & \multicolumn{2}{c}{YOOCHOOSE} & \multicolumn{2}{c}{DIGINETICA} \\
			\midrule
			\#  clicks            & 557,248& & 982,961 & \\
			\#  training sessions & 369,859& & 719,470 & \\
			\#  test sessions     &  55,898& &  60,858 & \\
			\#  items             &  16,766& &  43,097 & \\
			Average session length    &    6&.16 &    5&.12 \\
			\bottomrule
		\end{tabular}
	}
	\label{Table1}
\end{table}
The dimensions of the item embeddings and attention are set to $d=200$ and $l=100$, respectively. 
We use the Adam optimizer with an initial learning rate $10^{-3}$ and a decay factor 0.1 for every 3 epochs. 
The batch size is set to 128 and $L2$ regularization is applied to avoid overfitting by setting $L2=10^{-5}$.

\smallskip\noindent\emph{Evaluation metrics.}
Like \citep{KDD18/STAMP,SIGIR19/CSRM}, we evaluate SR-IEM and the baselines using Recall@$N$ and MRR@$N$; we set $N$ to 20 in our experiments.

\section{Results and Discussion}
\label{Results and Discussion}

\subsection{Overall performance}
\label{Overall performance}
To answer RQ1, we compare SR-IEM to baselines in terms of Recall@20 and MRR@20.
The results are presented in Table~\ref{Table2}.
\begin{table}[t]
	\centering
	\caption{Model performance. The results of the best baseline and the best performer in each column are underlined and boldfaced, respectively. $^{\blacktriangle}$ denotes a significant improvement of SR-IEM over the best baseline using a paired $t$-test (p < 0.01).}
	\setlength{\tabcolsep}{0.022\linewidth}{
		\begin{tabular}{l cc cc}
			\toprule
			\multirow{2}{*}{Method}&
			\multicolumn{2}{c}{YOOCHOOSE}&
			\multicolumn{2}{c}{DIGINETICA}\\
			\cmidrule(lr){2-3} \cmidrule(lr){4-5}
			&Recall@20&MRR@20 &Recall@20&MRR@20\\
			\midrule
			S-POP    &30.44\duwen &18.35\duwen &21.06\duwen &13.68\\
			Item-KNN &51.60\duwen &21.81\duwen &35.75\duwen &11.57\\
			FPMC     &45.62\duwen &15.01\duwen &31.55\duwen &\phantom{0}8.92\\
			\midrule
			GRU4REC  &60.64\duwen &22.89\duwen &29.45\duwen &\phantom{0}8.33\\
			NARM     &68.32\duwen &28.63\duwen &49.70\duwen &16.17\\
			STAMP    &68.74\duwen &29.67\duwen &45.64\duwen &14.32\\
			CSRM     &69.85\duwen &29.71\duwen
			&\underline{51.69}\duwen    &16.92\\
			SR-GNN   &\underline{70.57}\duwen
			&\underline{30.94}\duwen
			&50.73\duwen &\underline{17.59}\\
			\midrule
			\textbf{SR-IEM }  &\textbf{71.15}$^{\blacktriangle}$
			&\textbf{31.71}$^{\blacktriangle}$
			&\textbf{52.35}$^{\blacktriangle}$
			&\textbf{17.64}\\
			\bottomrule
	\end{tabular}}
	\label{Table2}
	\vspace{-5pt}
\end{table}
First of all, for the baselines, we see that the neural models generally outperform the traditional methods. 
For instance, SR-GNN performs best on YOOCHOOSE in terms of both metrics while it loses against CSRM on DIGINETICA in terms of Recall@20. 
SR-GNN is able to explore complex transitions of items to generate accurate user preferences by applying a {GGNN} while CSRM incorporates neighbor sessions on the basis of NARM, leading to better performance than other baselines. 
Thus, we choose CSRM and SR-GNN for comparisons in later experiments. 

Next, we zoom in on the performance of SR-IEM. 
In general, SR-IEM outperforms all baselines on both datasets in terms of both metrics. 
For instance on YOOCHOOSE, SR-IEM shows a 2.49\% improvement in terms of MRR@20 against the best baseline SR-GNN, which is higher than the corresponding 0.82\% improvement in terms of Recall@20.
In contrast, on DIGINETICA, the corresponding improvement in terms of Recall@20 is relatively higher than MRR@20. This may be attributed to the size of item set.
Thus, SR-IEM is able to boost the ranking of target items for the cases with relatively few candidate items while it is even more effective on hitting the target item for cases with relatively many candidate items.

In addition, we analyze the computational complexity of SR-IEM as well as the best  baselines, i.e., CSRM and SR-GNN. 
For CSRM and SR-GNN, the computational complexity is $O(td^2+dM+d^2)$ and $O(s(td^2+t^3)+d^2)$, respectively, where $t$ denotes the session length and $d$ is the dimension of item embeddings. 
Here, $M$ is the number of incorporated neighbor sessions in CSRM and $s$ is the number of training steps in GGNN. 
For SR-IEM, the computational complexity is $O(t^2d+d^2)$, which mainly comes from the importance extraction module $O(t^2d+d^2)$ and
from the other components $O(d^2)$. 
As $t < d$ and $d\ll M$ \citep{SIGIR19/CSRM}, the complexity of SR-IEM is clearly lower than that of SR-GNN and CSRM. 
To confirm this empirically, we present the training and test times of SR-IEM as well as CSRM and SR-GNN in Table~\ref{Table3}.
\begin{table}[h]
	\centering
	\caption{Computational complexity and efficiency. We set the training and test time of SR-IEM to 1 unit, respectively. {Then, we can find the relative time cost of each corresponding model against SR-IEM.}}
	\setlength{\tabcolsep}{0.015\linewidth}{
		\begin{tabular}{llrrrr}
			\toprule
			\multirow{2}{*}{Method}&
			\multirow{2}{*}{Complexity}&
			\multicolumn{2}{c}{YOOCHOOSE}&
			\multicolumn{2}{c}{DIGINETICA}\\
			\cmidrule(lr){3-4} \cmidrule(lr){5-6}
			&&training&test&training&test\\
			\midrule
			%
			CSRM   &$O(td^2+dM+d^2)$     & 4.91& 18.62& 4.63& 19.32 \\
			SR-GNN &$O(s(td^2+t^3)+d^2)$ & 3.12&  2.89& 2.56&  2.75 \\
			SR-IEM &$O(t^2d+d^2)$        & 1.00&  1.00& 1.00&  1.00 \\
			\bottomrule
	\end{tabular}}
	\label{Table3}
	\vspace{-3pt}
\end{table}
\vspace{-1pt}
We find that SR-IEM has clearly lower time costs than CSRM and SR-GNN.
This indicates that compared to the baselines, SR-IEM can perform best in terms of both recommendation accuracy and computational complexity, making it practicable for potential applications. 

\vspace{-8pt}
\subsection{Impact of session length}
\label{Impact of session length}
\begin{figure}[t]
	\centering
	\begin{minipage}[ht]{0.48\columnwidth}
		\includegraphics[clip,trim=0mm -2mm 22mm 18mm,width=\columnwidth]{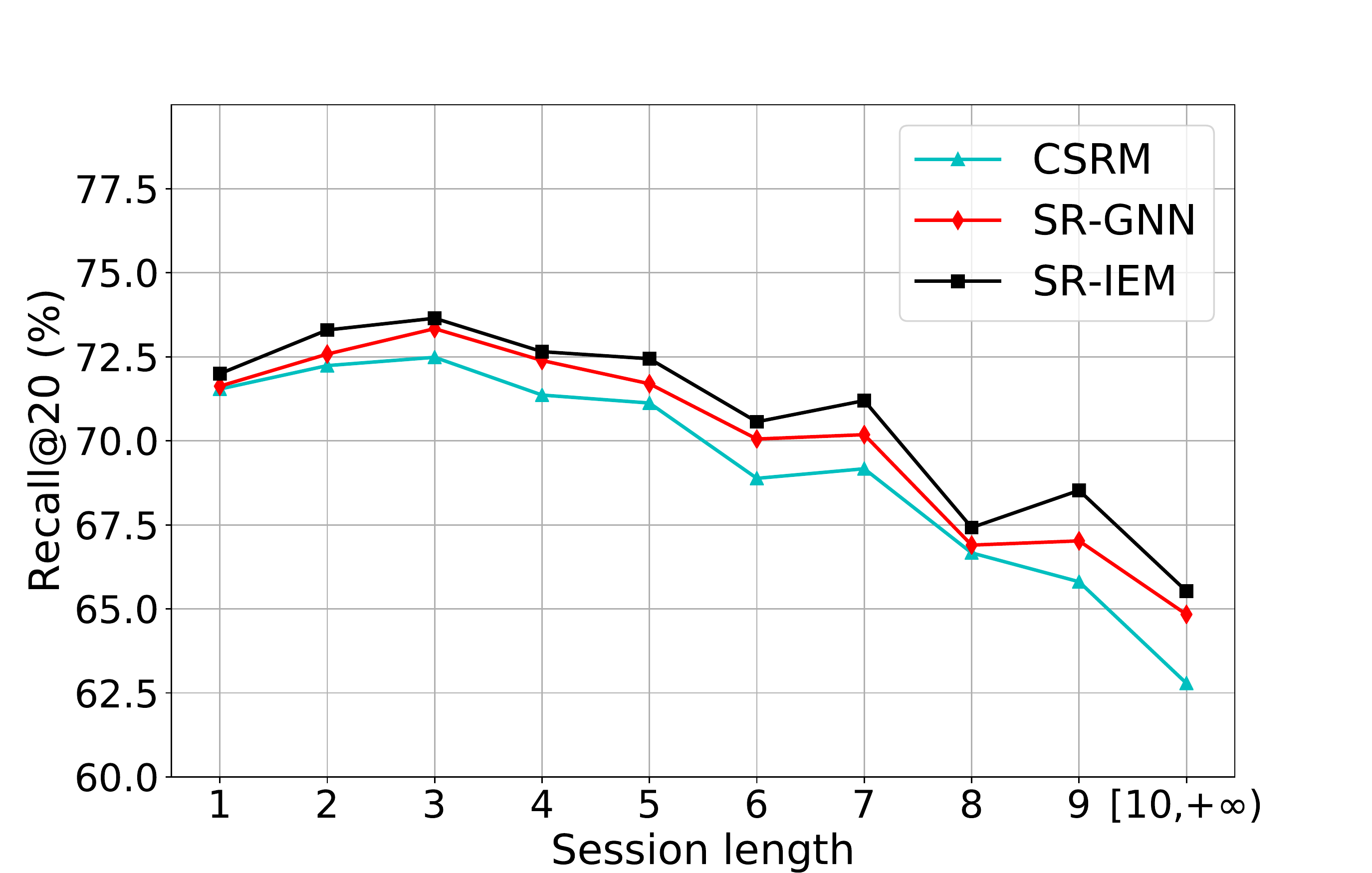}
		\vspace{-8pt}
		\subcaption{\textbf{Recall@20 on YOOCHOOSE.}}
		\label{Figure2.1}
	\end{minipage}
	\hspace{0.5pt}
	\begin{minipage}[ht]{0.48\columnwidth}
		\includegraphics[clip,trim=0mm -2mm 22mm 18mm,width=\columnwidth]{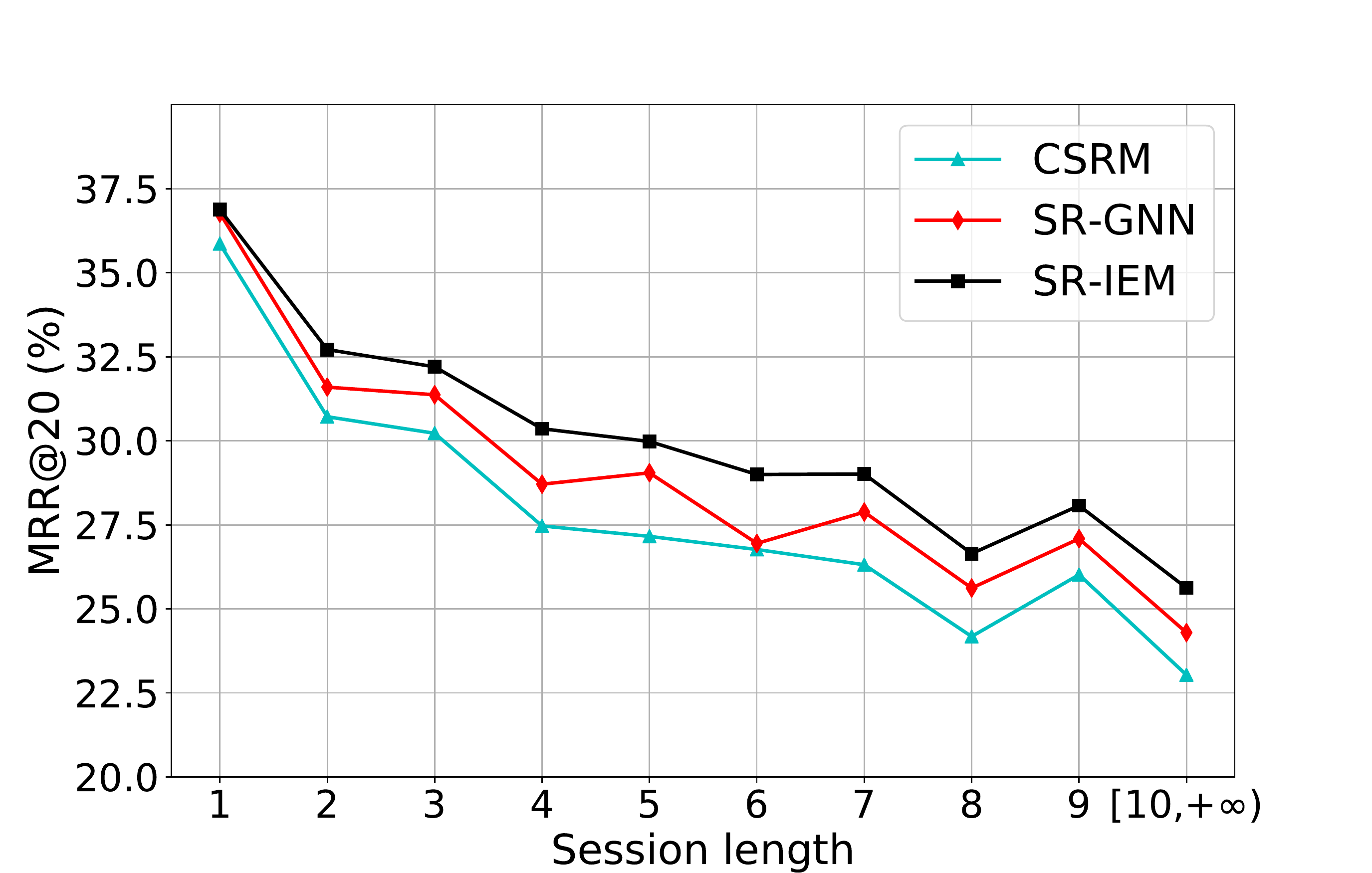}
		\vspace{-8pt}
		\subcaption{\textbf{MRR@20 on YOOCHOOSE.}}
		\label{Figure2.2}
	\end{minipage}
	\begin{minipage}[ht]{0.48\columnwidth}
		\includegraphics[clip,trim=0mm -2mm 22mm 18mm,width=\columnwidth]{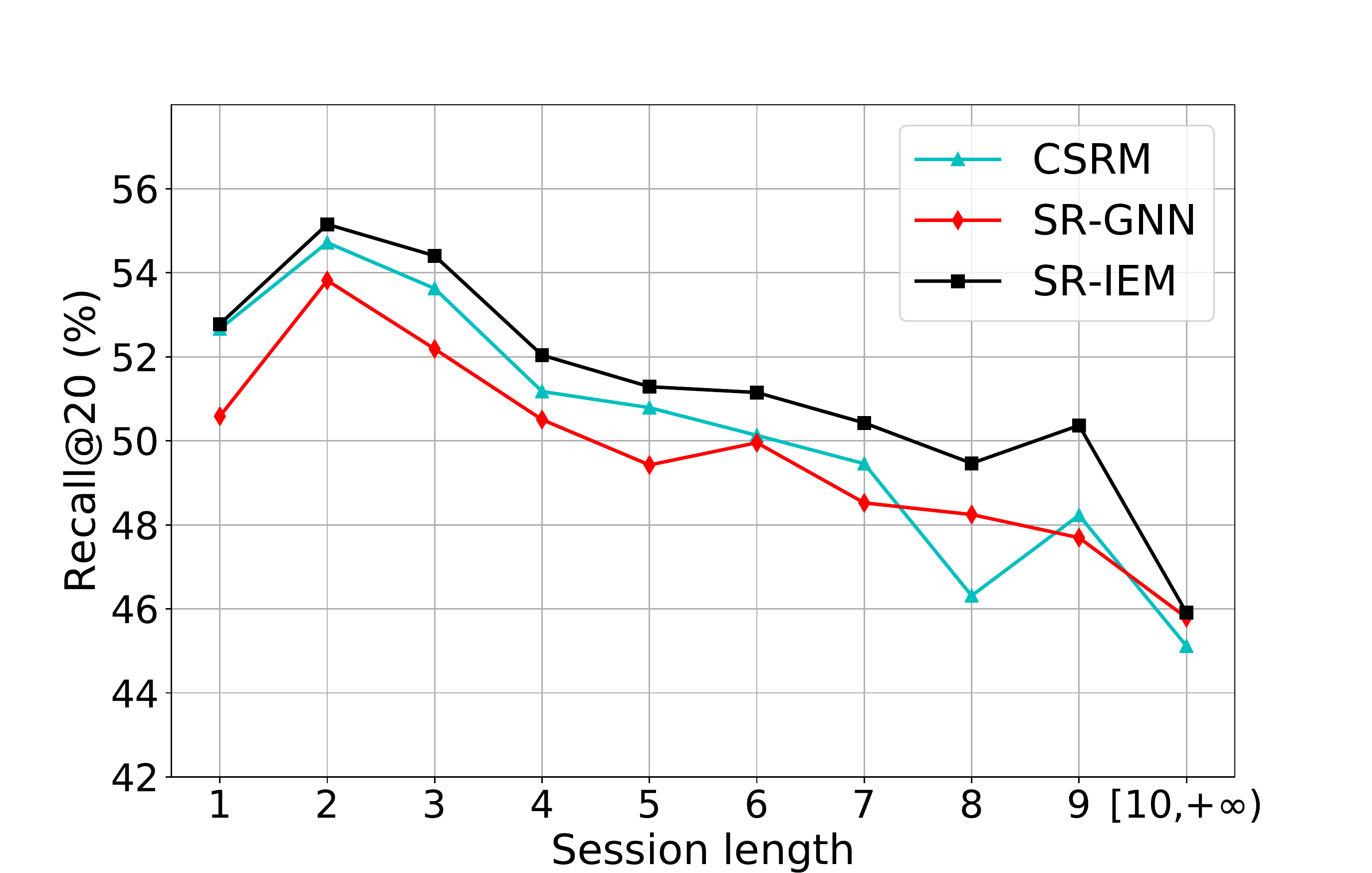}
		\vspace{-8pt}
		\subcaption{\textbf{Recall@20 on DIGINETICA.}}
		\label{Figure2.3}
	\end{minipage}
	\hspace{0.5pt}
	\begin{minipage}[ht]{0.48\columnwidth}
		\includegraphics[clip,trim=0mm -2mm 22mm 18mm,width=\columnwidth]{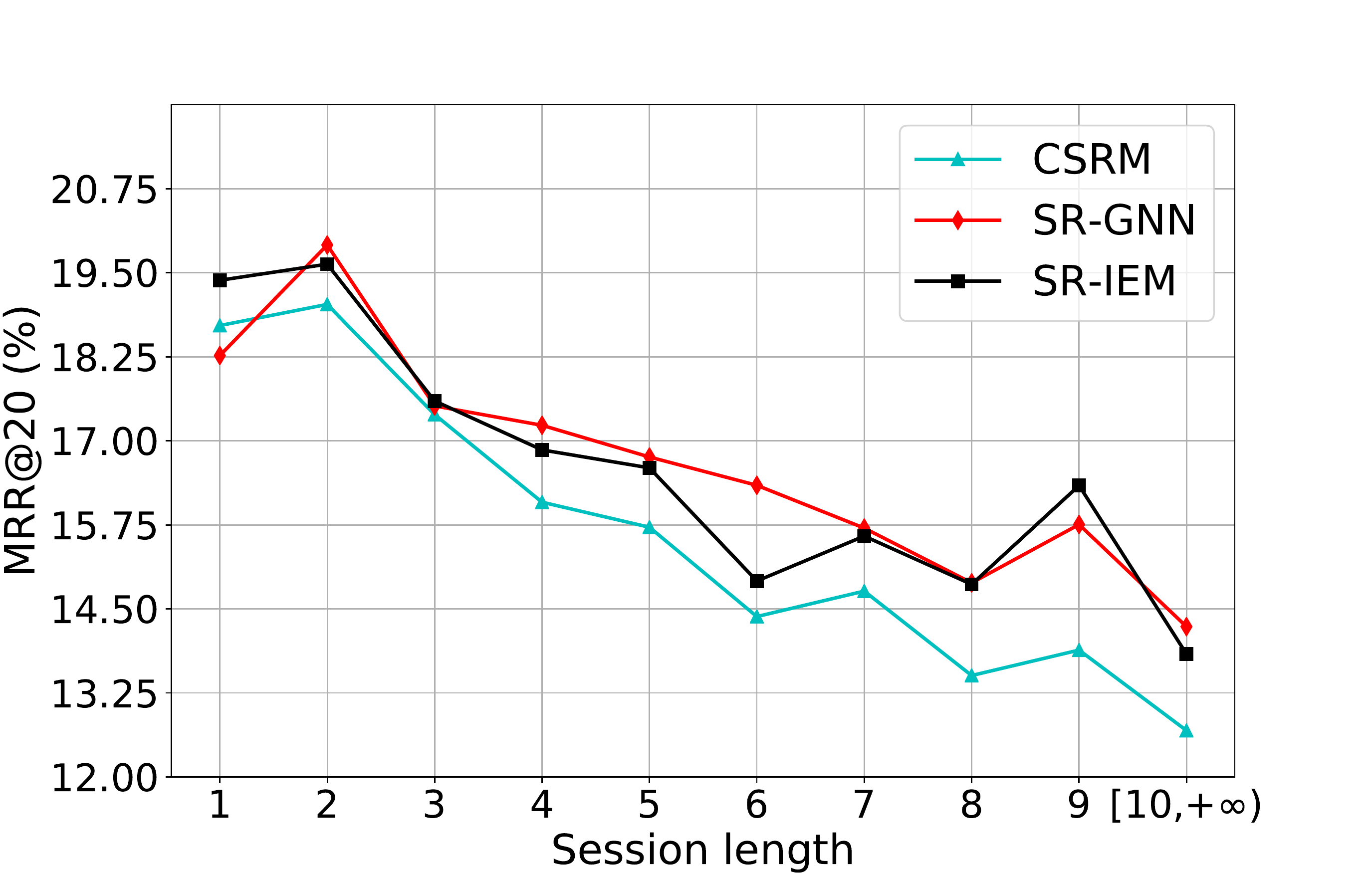}
		\vspace{-8pt}
		\subcaption{\textbf{MRR@20 on DIGINETICA.}}
		\label{Figure2.4}
	\end{minipage}
	\caption{Model performance under varying session lengths.}
	\label{Figure2}
	\vspace{-5pt}
\end{figure}

To answer RQ2, we plot the results of SR-IEM, CSRM and SR-GNN in terms of Recall@20 and MRR@20 in Fig.~\ref{Figure2}.
As for Recall@20, we see that as the session length increases, the performance of the three models on YOOCHOOSE and DIGINETICA increases first and then shows a continuous downward trend. 
The improvement of SR-IEM over CSRM and SR-GNN is more obvious for sessions lengths 4--7 than for lengths 1--3. 
When the session length is too short, IEM is not able to distinguish the item importance very well. 
As the length increases, the effectiveness of IEM goes up. 

For MRR@20, all models display a consistent downward trend on YOOCHOOSE and DIGINETICA as the session length increases.
SR-IEM outperforms CSRM and SR-GNN at all lengths on YOOCHOOSE while losing to SR-GNN for some cases on DIGINETICA, e.g., at lengths 4 and 5. 
In addition, on DIGINETICA, the MRR@20 scores show a sharper decrease than the Recall@20 scores. 
The difference in Recall@20 and MRR@20 scores on the two datasets may be due to the fact that non-relevant items 
have a bigger negative impact on MRR@20 than on Recall@20.

\vspace{-5pt}
\subsection{{Analysis on} importance extraction module}
\label{Impact of importance extraction module}

\begin{figure}[t]
	\centering
	\begin{minipage}[ht]{0.48\columnwidth}
		\includegraphics[clip,trim=4mm 0mm 16mm 14mm,width=\columnwidth]{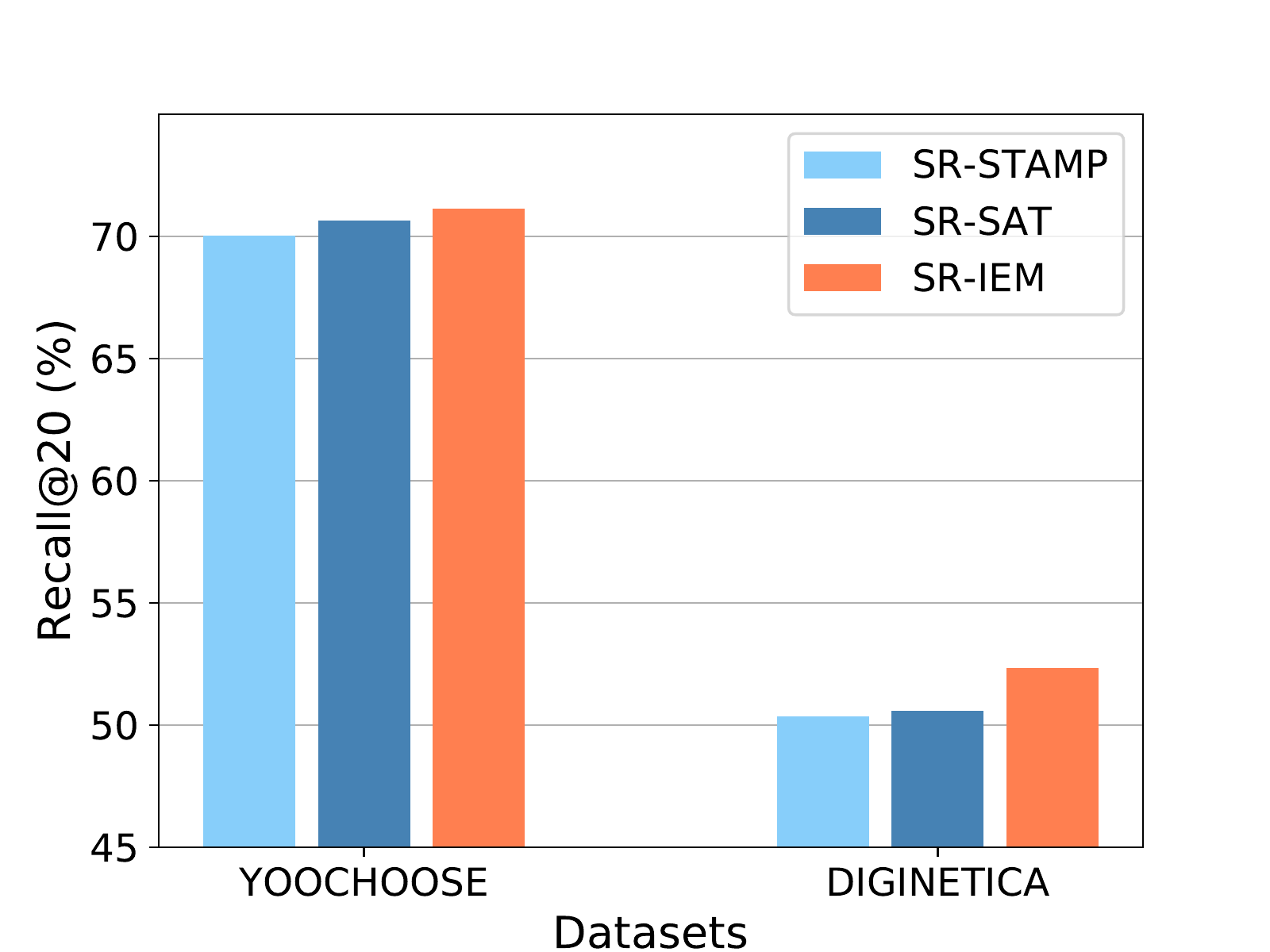}
		\vspace{-5pt}
		
		\subcaption{\textbf{Recall@20.}}
		\label{Figure3.1}
	\end{minipage}
	\begin{minipage}[ht]{0.48\columnwidth}
		\includegraphics[clip,trim=4mm 0mm 16mm 14mm,width=\columnwidth]{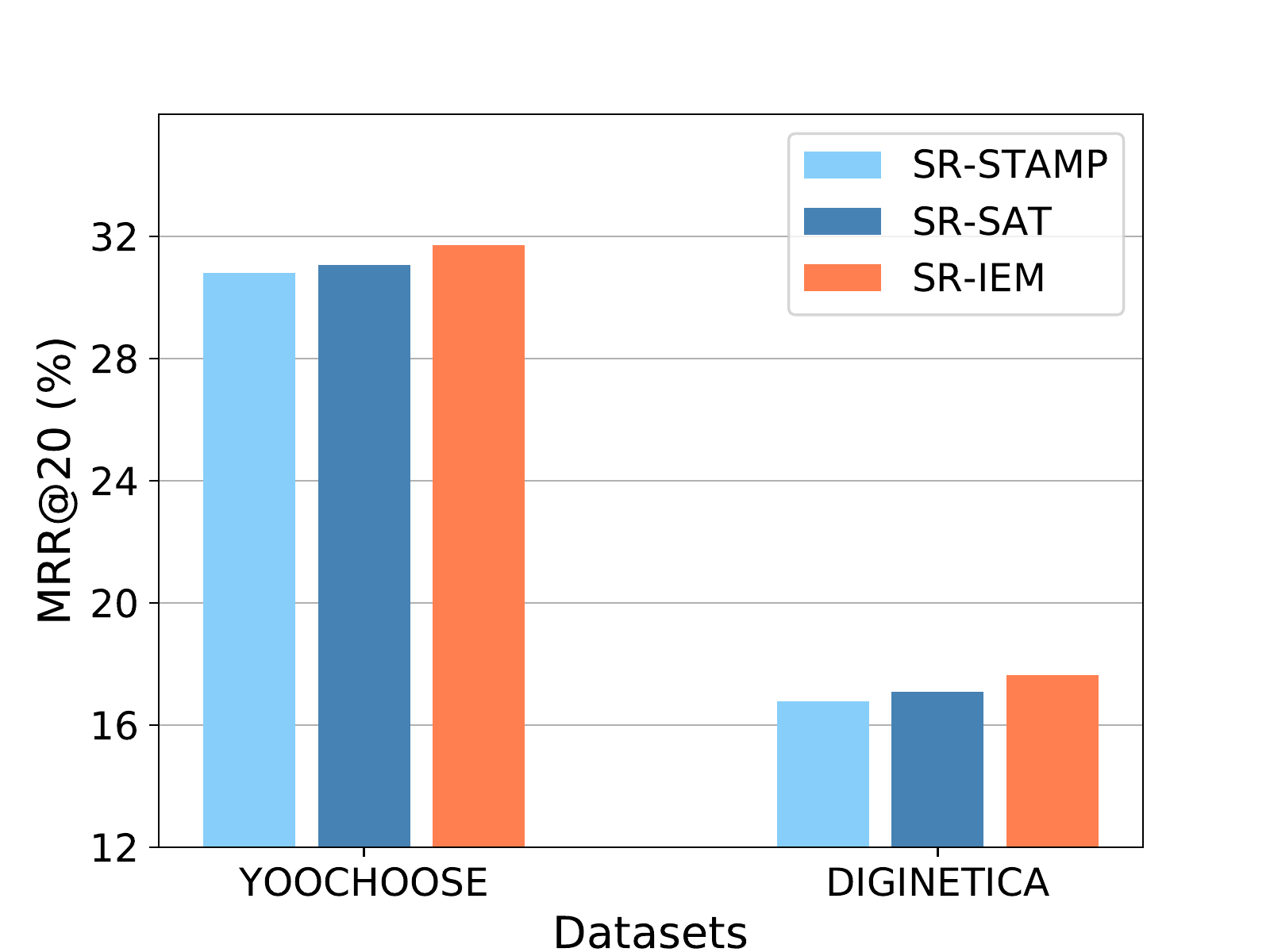}
		\vspace{-5pt}
		
		\subcaption{\textbf{MRR@20.}}
		\label{Figure3.2}
	\end{minipage}
	\caption{Model performance of various importance extraction methods used in our framework on two datasets. 
	}
	\label{Figure3}
	\vspace{-5pt}
\end{figure}

To answer RQ3, we replace IEM in SR-IEM with two alternatives and make comparison. 
We denote the variant models as
\begin{enumerate*}
	\item SR-STAMP: replace IEM with an attention mechanism proposed by \citep{KDD18/STAMP}; 
	here the mixture of all items and the last item in session is deemed as ``query.''
	\item SR-SAT: utilize a self-attention mechanism~\citep{NIPS17/Attention} to distinguish the item importance, and then aggregate them using an average pooling strategy \citep{Arxiv18/AttRec}. 
\end{enumerate*}
The results are shown in Fig.~\ref{Figure3}.

\vspace{-1pt}

In general, SR-IEM achieves the best performance in terms of Recall@20 and MRR@20 on both datasets.
{SR-SAT} outperforms {SR-STAMP}. 
This could be due to the fact that SR-SAT considers the item-item relationship in a session by modeling the contextual signal, which helps to capture user's preference for generating correct item recommendations. 
However, SR-STAMP only takes a mixture of all items and the last item to determine the item importance, thus failing to accurately represent user's preference. 
In addition, it is difficult for both SR-SAT and SR-STAMP to eliminate non-relevant items in a session, which results in a negative effect on the recommendation performance. 
In contrast, the proposed IEM module can effectively locate important items and assign a relatively high weight to them for user preference modeling, in a way that avoids being distracted by other items in the session.

\section{Conclusions and Future Work}
\label{Conclusions and Future Work}

We have proposed an Importance Extraction Module for Session-based Recommendation (SR-IEM), that incorporates a user's long-term preference and his current interest for item recommendation. 
A modified self-attention mechanism is applied to estimate item importance in a session for modeling a user's long-term preference, which is combined with user's current interest indicated by the last item to produce the final item recommendations. 
Experimental results show that SR-IEM achieves considerable improvements in terms of Recall and MRR over state-of-the-art models with reduced computational costs compared to competitive neural models. 
As to future work, 
we would like to apply the Importance Extraction Module to other tasks, e.g., dialogue systems and conversation recommendation.

\bibliographystyle{ACM-Ref-Format}
\bibliography{reference}

\end{document}